\documentclass[12pt]{iopart}

\usepackage[utf8]{inputenc}
\usepackage{graphicx}
\usepackage{dcolumn}

\begin{document}
\title{Directional detection of dark matter with diamond}
\author{Mason C. Marshall$^{1,2}$, Matthew J. Turner$^{3,4}$, Mark J.H. Ku$^{3,5,6,7}$, David F. Phillips$^2$, Ronald L. Walsworth$^{1,2,7,8}$}
\address{$^1$ Department of Electrical and Computer Engineering, University of Maryland, College Park, Maryland, 20742, USA}
\address{$^2$ Center for Astrophysics $\vert$ Harvard \& Smithsonian, Cambridge, Massachusetts, 02138, USA}
\address{$^3$ Department of Physics, Harvard University, Cambridge, Massachusetts, 02138, USA}
\address{$^4$ Center for Brain Science, Harvard University, Cambridge, Massachusetts, 02138, USA}
\address{$^5$ Department of Physics and Astronomy, University of Delaware, Newark, Delaware, 19716, USA}
\address{$^6$ Department of Materials Science and Engineering, University of Delaware, Newark, Delaware, 19716, USA}
\address{$^7$ Quantum Technology Center, University of Maryland, College Park, Maryland, 20742, USA}
\address{$^8$ Department of Physics, University of Maryland, College Park, Maryland, 20742, USA}
\date{draft: \today}
\vspace{10pt}
\begin{indented}
\item[]September 2020
\end{indented}

\begin{abstract}
    Searches for WIMP dark matter will in the near future be sensitive to solar neutrinos.  Directional detection offers a method to reject solar neutrinos and improve WIMP searches, but reaching that sensitivity with existing directional detectors poses challenges.  We propose a combined atomic/particle physics approach using a large-volume diamond detector.  WIMP candidate events trigger a particle detector, after which spectroscopy of nitrogen vacancy centers reads out the direction of the incoming particle.  We discuss the current state of technologies required to realize directional detection in diamond and present a path towards a detector with sensitivity below the neutrino floor.
\end{abstract}

\section{Introduction}

Weakly interacting massive particles (WIMPs) remain one of the most compelling dark matter (DM) models after decades of searches \cite{WIMPCandidatesReview2018}.  WIMPs are a dark matter candidate which, via thermal freeze-out, arise in an abundance consistent with the observed dark matter density.  Additionally, WIMP candidates arise naturally in many beyond-standard-model theories \cite{WIMPWaningReview2018}.  Direct detection of WIMPs is a long-term and ongoing effort, with exclusion bounds on cross-sections and WIMP masses shrinking by several orders of magnitude each over the past three decades \cite{Schumann2019}.

 Upcoming detectors are expected to reach sensitivity levels allowing detection of solar neutrinos \cite{NeutrinoFloor,AtmosNeutrinos}.  Coherent elastic neutrino scattering events induce single nuclear recoils \cite{Akimov1123}, and therefore pass cuts looking for electron recoils or multiple scattering due to backgrounds such as cosmic rays and radioisotope contamination of target materials and instrumentation.  Solar neutrinos thus constitute an irreducible background for future WIMP detectors -- the ``neutrino floor.''  Discovering a dark matter particle will therefore require identifying annual modulation of a WIMP signal as the Earth's velocity changes relative to a putative WIMP ``wind'' \cite{DMannualmodulation2013}. To do so on top of a large solar neutrino background will demand dozens of DM events.

An alternative approach to extend WIMP searches below the neutrino floor is directional detection \cite{ReadoutDirectional,DirBeyondNeutrino}.  If the direction of an incoming particle can be determined, solar neutrino events can be discriminated, allowing a low-background search for cosmological dark matter particles.   A number of routes are being pursued towards directional DM detection using a variety of target materials, each with its own advantages and disadvantages.  Gas-phase detectors \cite{CYGNUS2020,NEWAGE2010,DRIFT2015,MIMACRadon,DMTPC2017,ColumnarAngularXe} have well-developed directional discrimination; but because the sensitivity of a DM detector depends on its total mass, gas-phase detectors will require extremely large volumes to reach the sensitivity of liquid- or solid-phase detectors.   Emulsion-phase detectors \cite{NEWSDMpotential,NEWSDM2020} have higher target density than gaseous detectors, but require lengthy readout times post-exposure; and the time-integrated directional detection they provide necessitates substantially higher exposures to probe below the neutrino floor \cite{Time-integratedDD}.  Directional scintillators \cite{ADAMOcrystals} reach solid-state target density and use traditional particle physics readout techniques, but do not give directional information for individual events --- the measured energy for each event is shifted depending on the incoming particle's direction, requiring day-night variation (and thus many events) to extract a directional signal.  Dual-phase liquid noble gas detectors may be extended to partial directional discrimination by taking advantage of columnar recombination \cite{DirectionalAr}, but extensive development is still required.  

We envision a ``hybrid'' approach whereby directional detection can be incorporated into a more traditional WIMP detector, with a solid-state target mass and using well-developed instrumentation and background-discrimination methods~\cite{proposal}.  Such a hybrid detector would identify WIMP candidate events using traditional methods like charge, phonon, or scintillation collection in a solid-state, semiconductor detector.  A WIMP recoil in such a detector would impart substantial kinetic energy to a target nucleus, initiating a chain of secondary nuclear recoils; this would leave behind a characteristic ``track'' of crystal damage, comprising lattice vacancies, interstitial nuclei, and distorted bonds.  These damage tracks are predicted to exhibit measurable head-tail asymmetry, and to be oriented near the direction of the incoming particle \cite{proposal}.  In the hybrid detector, a WIMP candidate event in a segmented target would trigger traditional particle-detection systems.  The relevant segment would then be removed from the bulk of the detector; the crystal damage from the WIMP would be measured and the direction read out.  Together with the detector position at the time of the event, the damage track’s direction gives the WIMP incident direction relative to the sun and the galactic halo.

For this hybrid detection method, we propose to develop a detector with diamond as its target material.  Diamond is the subject of active study for next-generation semiconductor-based WIMP detectors --- it offers excellent semiconductor properties, and the relatively light carbon nucleus provides an improved sensitivity profile for low-mass DM candidates \cite{Kurinsky2019} compared to xenon or other heavy target materials. We will not provide a detailed discussion of traditional event detection in diamond here, but we note that this is an active topic of research, with ongoing investigations around the world \cite{Kurinsky2019,DiamondCryo2020}.  To achieve sensitivity below the neutrino floor, a large total target mass will be required --- comparable to the multi-ton masses of other next-generation WIMP detectors, with diamond volumes at the cubic meter scale.  Modern diamond growth techniques using chemical vapor deposition (CVD) enable repeatable, fast growth of uniform crystals, and with realistic development should be capable of producing the volumes of diamond necessary for such a detector.  Optimized crystal growth protocols \cite{PurpleDiamond} lead to both advantageous semiconductor properties, allowing for very sensitive charge and phonon extraction, and also very uniform, low crystal strain.

While our ``hybrid'' approach takes advantage of traditional detector infrastructure combined with interrogation of solid state defects, complementary proposals for crystal defect-based WIMP searches are explored in the literature.  Natural minerals could be used as ``paleodetectors'' \cite{Paleo2019n1,Paleo2020}, with defect spectroscopy revealing WIMP-induced damage tracks accumulated over geological timescales; this method would yield large exposures with modest target masses, but because the signal is integrated over a long exposure, directional information is difficult to extract \cite{Paleo2019n2}.  Additionally, real-time monitoring of the creation of single color centers has been proposed as a method to search for sub-GeV dark matter \cite{ColorCenterMonitoring2018}; this proposal is sensitive to lower-mass DM particles that cannot induce damage tracks, but continuous optical monitoring makes obtaining a large exposure more difficult than with a hybrid detector.

\begin{figure}[htbp!]
\begin{center}
\includegraphics*[width=\columnwidth]{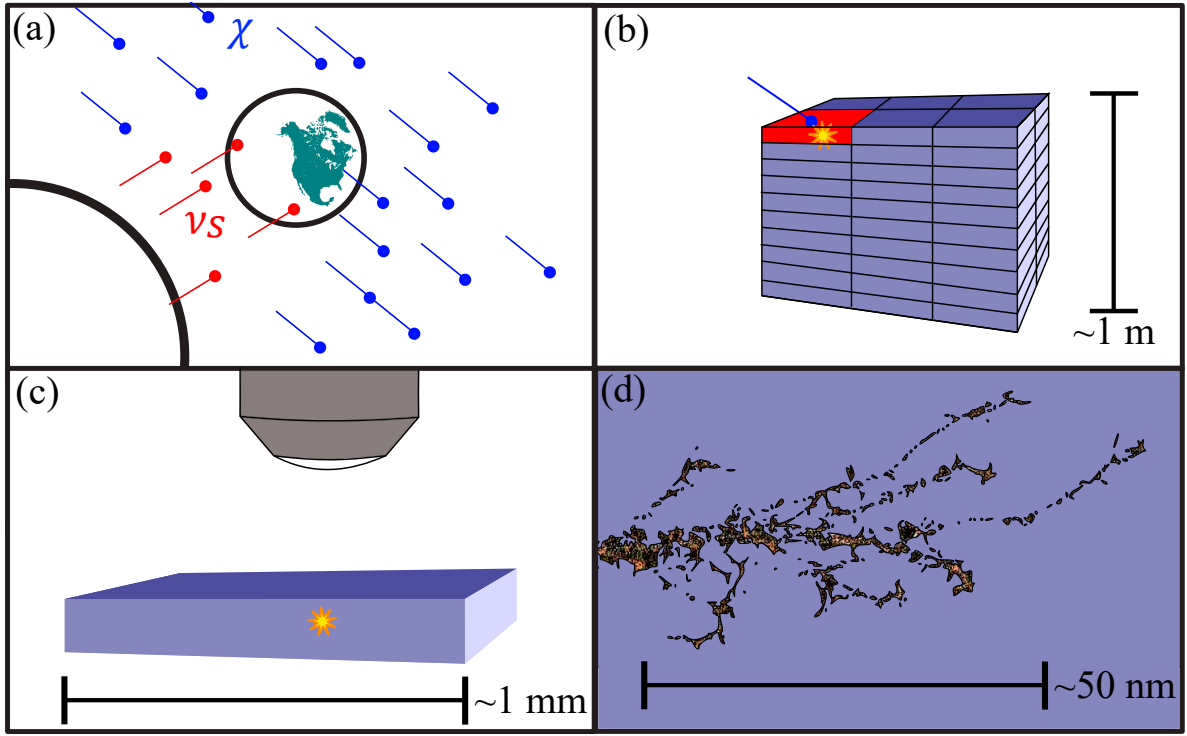}
\caption{Schematic overview of the proposed dark matter detection procedure.  \textbf{(a)} The earth passes through the galactic WIMP ($\chi$) distribution (WIMP ``wind'') as solar neutrinos ($\nu_s$) arrive from the sun.  An underground detector is carried by the rotation of the earth and interacts with the incoming neutrinos and WIMPS.  \textbf{(b)} Occasionally, a WIMP or neutrino initiates a nuclear recoil cascade via collision with a carbon nucleus in the detector, depositing energy and damaging the crystal lattice; the event is detected by charge, photon, or phonon collection.  The time of the event is recorded, giving the orientation of the detector relative to the Sun and galactic WIMP wind.  \textbf{(c)} The triggered segment is removed from the detector bulk; using diffraction-limited optical methods, the nuclear recoil track is localized to a $\sim\mu$m$^3$ volume.  \textbf{(d)} Finally, the distribution of crystal damage in the nuclear recoil track is measured and reveals the incoming direction of the particle, allowing statistical discrimination of WIMPs and neutrinos. }
\label{fig:DMoverview}
\end{center}
\end{figure}

For recoils with energy in the range 10--100 keV, corresponding to WIMP masses in the 1--100 GeV range, SRIM simulations predict damage tracks will be tens of nanometers long \cite{proposal}.  After localizing a WIMP candidate event to a single CVD diamond chip \cite{LZconf2016}, of order 1 mm$^3$, within the cubic meter bulk of the detector, key technical challenges remain to extracting the direction of the incoming particle.  Scanning the entire chip with resolution below 10 nm would be time-prohibitive, and wide-field techniques are diffraction-limited by the wavelength of the applied laser light ($\sim$0.5 $\mu$m).  We therefore propose a three-step process for a diamond directional detector, illustrated in Fig.\ \ref{fig:DMoverview}:\\

\begin{itemize}
\item First, an event is registered and triangulated on the mm scale by photon, phonon, or charge-carrier collection.
\item Second, using diffraction-limited optics the event’s damage track is localized to a $\sim\mu$m voxel within the diamond.
\item Finally, superresolution optical methods or high-resolution x-rays map the damage track at the nanoscale, yielding the required directional information.
\end{itemize}

In the following sections, we describe our initial work on track localization and super-resolution imaging to determine the direction from which a WIMP or neutrino arrives at a DM detector.  We additionally provide estimates of the sensitivity and measurement speeds that must be achieved for this detection method to be viable.  Sensitivity estimates are informed by the expected damage track length of O(10-100) nm and crystal damage of O(50-300) vacancies, both determined from simulations using the SRIM software package \cite{proposal,SRIM}.  Measurement speeds are benchmarked by comparison to the expected event rate from coherent solar neutrino scattering, largely of $^8$B solar neutrinos. We estimate this rate to be O(10-30) events per year for a m$^3$-scale detector, based on theoretical studies for target nuclei with similar masses to carbon \cite{Strigari2009,Papoulias2018} as well as the recent non-detection of solar neutrino scattering by the XENON1T experiment \cite{XENON1T2020}.  

\section{Recoil localization with strain mapping by NV spectroscopy}

The great advantage of a diamond detector, beyond its sensitivity to low-mass WIMP candidates, is the existence of well-characterized, photoactive quantum defects.  Using quantum defects as integrated sensors for local crystal damage is the key enabling technology for a directional DM detector in diamond.  Specifically, the nitrogen vacancy (NV) center has been extensively characterized and used for a wide variety of quantum sensing applications \cite{NVNanoReview2014,SingleNVMagSensing2008,SingleNVElecSensing2011,EnsembleMagImaging2011}.  In this section, we evaluate the use of NV centers to accomplish the mm-to-$\mu$m scale localization step for directional detection.

\begin{figure}[htbp!]
\begin{center}
    \includegraphics*[width=\columnwidth]{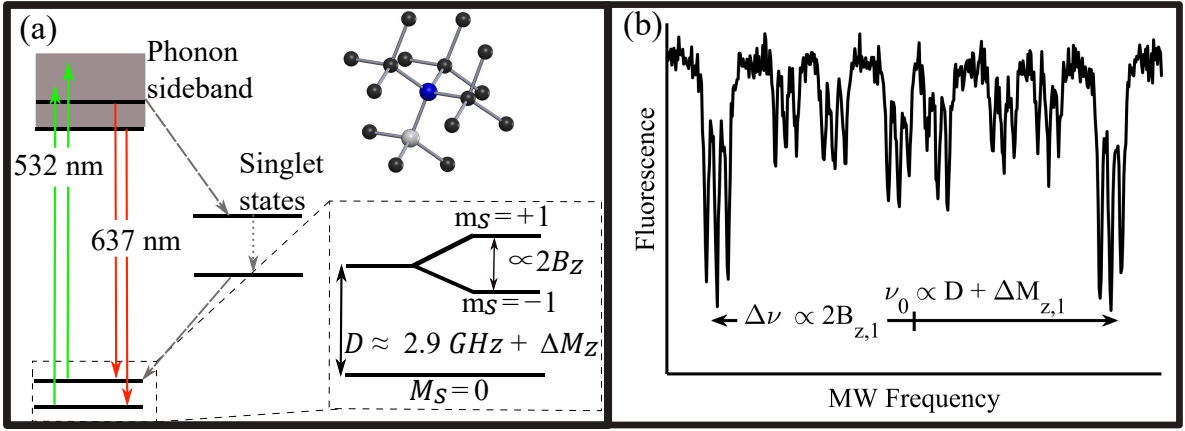}
\caption{\textbf{(a)} Negatively charged NV center energy level structure (S=1).  Electronic energy level transitions are excited into vibrational continuum states with green light and emit red light.  The m$_s=\pm$1 spin level exhibits reduced red fluorescence due to the presence of an alternate, non-radiative decay path.  Microwave spectroscopy enables strain and magnetic field measurements. (\emph{See text.})  Inset: NV diamond crystal structure. Here, carbon atoms of the diamond lattice are shown in black; the gray ball represents a vacancy in the lattice structure while the blue represents a nitrogen atom replacing one of the carbon atoms.  \textbf{(b)} Example NV spectrum  with an applied bias magnetic field.  Four pairs of hyperfine triplet resonances are produced by the four NV directions in an ensemble aligned with the crystal axes; each resonance has a center frequency $\nu_0$ which depends on temperature and local strain, as well as a frequency splitting $\Delta\nu$ between the m$_s$=0 to m$_s$=$\pm$1 states proportional to the magnetic field component along the NV axis, $B_{z,i}$ where $i=1,2,3,4$.}
\label{fig:NVlevels2}
\end{center}
\end{figure}

The NV center is a fluorescent point defect consisting of a substitutional nitrogen impurity adjacent to a lattice site vacancy, illustrated in Fig.~\ref{fig:NVlevels2}a.  The defect features an electron in a spin-1 ground state with a microwave transition frequency which is sensitive to magnetic and electric fields, as well as local crystal strain.  The spin state can be optically initialized, and read out via the spin-dependent red fluorescence emitted after excitation with green light.  Via their strain sensitivity, NV centers act as integrated sensors for nearby crystal damage \cite{StrainPaper}, presenting methods for both diffraction-limited and superresolution detection of nuclear recoil tracks.

Here, we consider the use of NV sensing to detect crystal strain features --- local deformations in the diamond crystal lattice.  Strain --- a relative change in the lattice spacing from its equilibrium value --- occurs in CVD diamond both due to imperfections which propagate during growth and flaws in surface processing \cite{Friel2009}.  Strain features in diamond have been studied using a variety of methods, including birefringence imaging \cite{BirefringenceStrain2014}, x-ray tomography \cite{E6xray2008}, and Raman imaging \cite{RamanStrain2011}.  Because of its deleterious effect on NV center-based magnetic field measurements \cite{SensitivityReport}, strain has been studied extensively in this regard \cite{StrainPaper,AusStrain2019}.  While intrinsic strain can be a contaminant to observing damage tracks left by particle scattering (see section \ref{sec:strainbackground}), the same methods used to study intrinsic strain can be used to pinpoint the position of WIMP-induced damage tracks.

\subsection{State of the art in NV strain imaging}
\label{sec:Stateoftheart}

\begin{figure}[htbp!]
\begin{center}
\includegraphics*[width=\columnwidth]{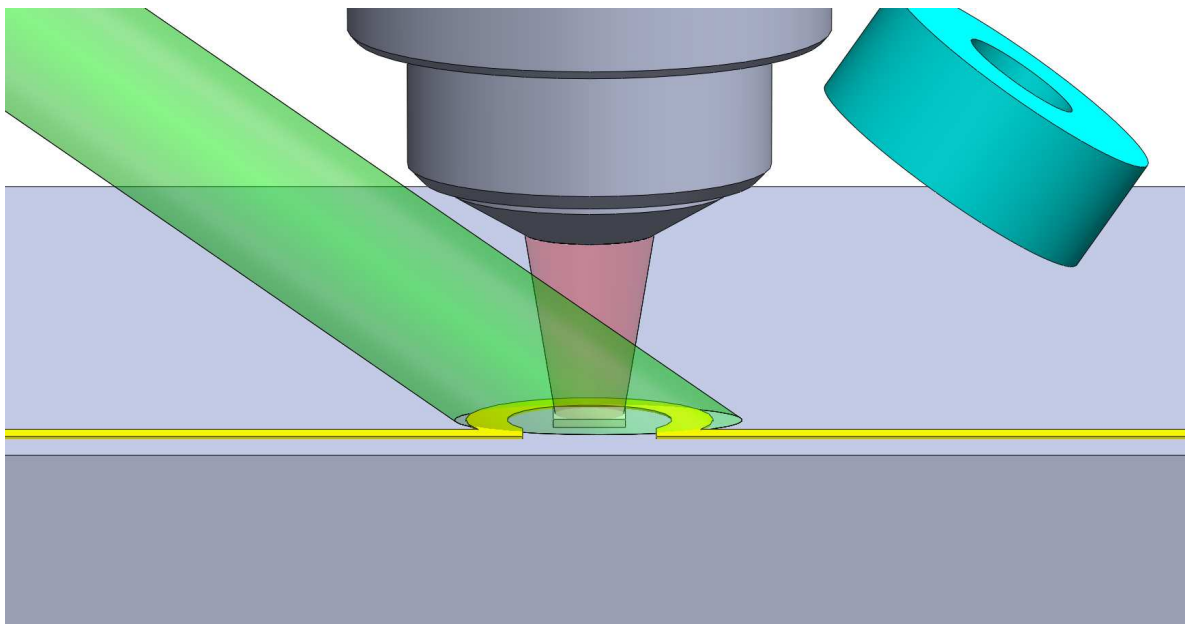}
\caption{Simplified schematic of a quantum diamond microscope (QDM), as used for wide-field strain imaging with nitrogen vacancy (NV) centers \cite{QDMreview}.  An NV-bearing diamond is illuminated by green laser light; NV centers emit red fluorescence, which is collected by a microscope objective and focused onto a camera (not shown).  The ground-state spin transition is probed by microwaves emitted by a waveguide (gold).  A bias field is provided by permanent magnets (teal) and/or Helmholtz coils (also not shown).}
\label{fig:qdm}
\end{center}
\end{figure}

NV strain imaging measurements are typically performed using a “quantum diamond microscope” (QDM) \cite{QDMreview}, as illustrated in Fig.\ \ref{fig:qdm}.  An NV-rich diamond is illuminated with green laser light, and the spin-dependent red fluorescence is collected by a microscope objective and imaged with a camera.  The NV spin is addressed by microwaves, and the transition frequencies determined by resonant reduction in the fluorescence (Fig.\ \ref{fig:NVlevels2}b).  Magnetic fields split the transition frequencies, while changes in strain and temperature yield common-mode shifts as shown in Fig.~\ref{fig:NVlevels2}.  Diamond is an excellent thermal conductor, and temperature thus can be assumed to be homogeneous across the diamond; spatially resolved measurements of common-mode frequency shifts are therefore maps of relative crystal strain.  This method has been used to study strain at boundaries in polycrystalline diamond \cite{EnglundStrain2016}; strain induced by implanted ions \cite{AusStrain2019}; and strain due to growth defects in CVD diamond layers \cite{StrainPaper}.  The strain sensitivity achievable with such measurements depends on the particular properties of the diamond samples used, such as density and coherence time of the NV centers used, as well as experimental factors such as camera sensitivity and resolution and instrumentation noise.  To evaluate the suitability of QDM strain measurement for damage track localization, we compare in Table \ref{table:Samples} the results reported in previous strain-imaging experiments.  We also extrapolate strain sensitivity from the reported sensitivity of some magnetic QDM experiments.
\begin{table}[h]
\begin{center}
\caption{Sample strain imaging measurements from the literature for diamonds with various nitrogen vacancy (NV) densities and layer thicknesses.  We converted the reported noise floors from each reference into units of relative strain.  * indicates magnetic-field imaging experiments from which we extrapolated an equivalent strain sensitivity.}
\label{table:Samples}
\begin{tabular}{||c|c|c|c|c|c||}
\hline
Sample type & Strain noise  & Lateral & Averaging & Ref. \\
& floor & resolution & time &\\
\hline\hline
Ensemble doped & 1$\times10^{-7}$	& 4 $\mu$m & 3 hours & \cite{StrainPaper} \\
\hline
Thin implant & 1$\times10^{-6}$ & 1 $\mu$m & 10 hours & \cite{AusStrain2019} \\
\hline
Single NVs & 4$\times10^{-6}$ & $<$1 $\mu$m & 120 seconds & \cite{EnglundStrain2016} \\
\hline
Ensemble doped* & 1$\times10^{-7}$ & 5 $\mu$m	& 11 hours & \cite{GlennGGG} \\
\hline
Thin implant* & 5$\times10^{-7}$ & $<$1 $\mu$m & (not listed) & \cite{GeneticStrategy}\\
\hline \hline
\end{tabular}
\end{center}

 \end{table}

\subsection{Estimate of WIMP-induced strain signal in QDM}

Following \cite{proposal}, we assume a fractional strain from a WIMP scattering event in diamond of approximately $\frac{\Delta x}{x}\sim10^{-6}$ at a distance of 30 nm from a single vacancy, leading to an NV resonance shift of about 30 kHz.  As strain falls off as $1/r^3$, the strain from an entire damage cluster can be found by adding the strain from each individual point defect \cite{Eshelby1954}.  To determine the average signal, we find the strain shift for NV centers at each point within an imaging voxel, yielding the lineshape a QDM would see at that voxel.  We treat the damage track as a cylinder with height and radius 30 nm, and assume the strain within that cylinder induces a uniform 10$^{-6}$ shift per vacancy; this avoids complications in calculating the specific distribution of damage within the region.  Outside the cylinder, we calculate the strain using a $1/r^3$ scaling with distance from the cylinder.

\begin{figure}[htbp!]
\begin{center}
\includegraphics*[width=\columnwidth]{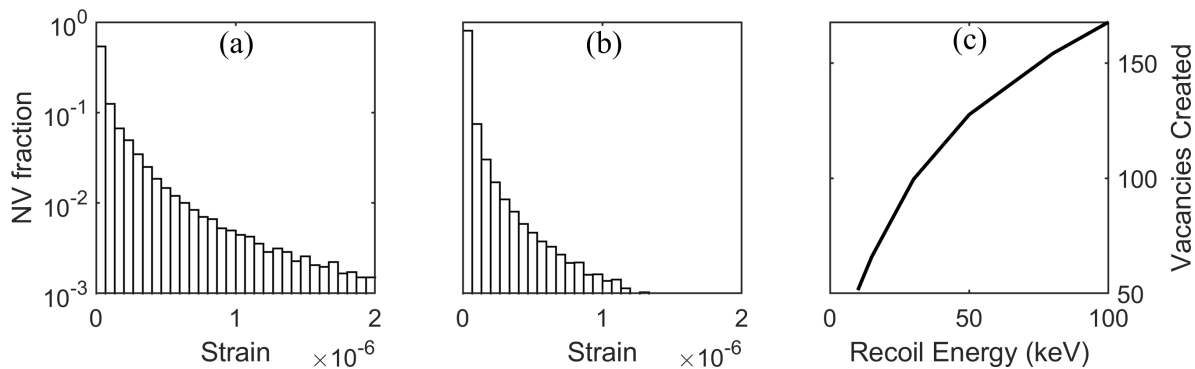}
\caption{\textbf{(a)} Proportion of NV centers experiencing a given strain, within a laterally diffraction-limited voxel of axial depth 1 $\mu$m and volume $\sim0.1\ \mu$m$^3$, containing a nuclear recoil track from 10 keV of deposited energy in a WIMP collision.  The mean signal from all such NV centers yields the voxel average properties given in Table \ref{table:strains}; details of the calculation are described in the text.  \textbf{(b)} As (a) but for a 1 $\mu$m$^3$ voxel.  \textbf{(c)} Number of vacancies created for different initial recoil energies, calculated using the SRIM simulation software \cite{SRIM}.}
\label{fig:qdmlw}
\end{center}
\end{figure}

An NV center is most sensitive to the projection of strain onto the axis between the nitrogen and the vacancy \cite{StrainPaper}.  Depending on the orientation of the crystal damage and its position within a voxel, the projection of strains onto different NV centers could add constructively, or average to zero.  We therefore extract two quantities from our estimate --- a mean of strain magnitudes, representing the case in which strain adds across the voxel; and a standard deviation, based on a distribution of strains with a mean of zero.  Figure \ref{fig:qdmlw} shows the distribution of strains within two different micron-thick voxels --- one with a diffraction-limited, 350 nm spot, and the other with a volume of 1 $\mu$m$^3$.  The mean of this distribution corresponds to the average strain shift in the case where strain adds constructively across the voxel; the standard deviation for the zero-mean case is extracted from a double-sided version of the same distribution, centered around zero.  These distributions are computed for a damage track from a 10 keV scattering event, yielding approximately 50 vacancies; the strain amplitudes for higher-energy events will increase accordingly.  Note that the horizontal scale is limited to 2$\times$10$^{-6}$ --- while NV centers near the damage track will experience strains up to $\sim$50$\times$10$^{-6}$, most NV centers in the voxel will have lower strains.  The mean strain magnitudes and linewidths can be extracted from the distributions in two ways - including or excluding these high-strain NVs.  The appropriate comparison will depend on experimental details, including (for example) the efficiency of the camera used in the QDM and its ability to detect light from a few, high-strain NVs, or the spin coherence properties of NVs in the diamond material.  Table \ref{table:strains} gives estimated voxel-averaged, WIMP-induced strain for different measurement parameters. These results vary by approximately one order of magnitude between $1\times10^{-7}$ and $3\times10^{-6}$.

\begin{table}[h]
\begin{center}
\caption{Estimated voxel-averaged strain signal amplitudes resulting from WIMP-induced damage tracks, for different experimental conditions as discussed in the text.  In comparison to Table \ref{table:Samples}, we note that the best QDM measurements to date have already achieved the needed strain sensitivities, albeit with long averaging times and/or insufficient spatial resolution (see Sec.~\ref{sec:improvingqdm}).}
\label{table:strains}
\begin{tabular}{||c|c|c|c|c|c||}
\hline
Signal type & Voxel size & High-strain NVs & Voxel-averaged strain\\
\hline\hline
Mean shift & Diffraction-limited & Included	& 7$\times$10$^{-7}$\\
\hline
Mean shift & Diffraction-limited & Excluded	& 3$\times$10$^{-7}$\\
\hline
Mean shift & 1 $\mu$m$^3$ & Included	& 2$\times$10$^{-7}$\\
\hline
Mean shift & 1 $\mu$m$^3$  & Excluded	& 1$\times$10$^{-7}$\\
\hline
St. Dev. & Diffraction-limited & Included	& 3$\times$10$^{-6}$\\
\hline
St. Dev.  & Diffraction-limited & Excluded	& 6$\times$10$^{-7}$\\
\hline
St. Dev.  & 1 $\mu$m$^3$ & Included	& 2$\times$10$^{-6}$\\
\hline
St. Dev.  & 1 $\mu$m$^3$  & Excluded	& 3$\times$10$^{-7}$\\

\hline \hline
\end{tabular}
\end{center}
 \end{table}

\subsection{Improving strain spectroscopy}
\label{sec:improvingqdm}

At $\sim1\times10^{-7}$, the noise floor of the best historical QDM measurements \cite{StrainPaper,QDMreview,GlennGGG} is below the estimated signal size for WIMP-induced damage tracks at the lowest end of the target sensitivity range, under all but the most pessimistic measurement scenarios summarized in Table \ref{table:strains}.  However, two technical barriers remain to WIMP track localization via strain spectroscopy.  First, the multi-hour averaging times required to resolve small strains are prohibitive, requiring hundreds of days to image 1000 $\mu$m-thick slices in a single detector segment.  (The noise floor of these measurements ties in here, as the 10$^{-7}$ limit is imposed by thermal fluctuations during the long averaging time).  Second, these techniques simultaneously collect fluorescence from the entire depth of the NV-bearing diamond layer, whereas to localize a damage track we must constrain its vertical position to a $\sim1\ \mu$m spot, likely by restricting the optical point spread function (PSF).  The reported spatial resolution for the most sensitive previous measurements compiled in Table \ref{table:Samples} is also worse than the requisite $\sim$1 micron; however, this resolution is limited by collection of light in a thick NV layer, so restricting the axial PSF would improve lateral resolution as well.

The acceptable time to localize a damage track must be faster than the expected event rate in the detector, to allow measurement of each event's direction as it is detected (rather than have the readout take longer than the exposure time).  Absent a high rate of WIMP events, this would be the $\sim$10-day expected period between neutrino floor events.  In order to account for a potential WIMP signal at the level of the neutrino floor, as well as the possibility of background events, we set a target of one to three days to localize each damage track to a um$^3$ voxel.  (An additional consideration could be the expected rate of background events, either in the detector proper or in the diamond chip after it is removed from the detector.  However, we assume a detector with sensitivity below the neutrino floor will have fewer background events than neutrino events.  If the QDM measurement is performed on-site at the shielded complex housing the detector, using a sample stage and optics made from low-background materials similar to those currently used in noble-gas-based WIMP searches \cite{XENONPMTS}, the rate of spurious damage tracks induced during measurement can be kept negligible.)

Advanced techniques from NV magnetic imaging offer a path to sufficiently fast measurement times.  All of the strain and magnetic field measurements in Table \ref{table:Samples} are performed using continuous-wave (CW) microwave excitation of the NV spins.  Measurement protocols using pulsed microwaves take advantage of the NV’s quantum coherence to dramatically improve data collection speed.  For example, wide-field magnetic field imaging measurements using a pulsed protocol recently achieved sensitivity equivalent to 1$\times10^{-7}$/$\mu$m$^{3/2}$/Hz$^{1/2}$ \cite{DQ4R}.  This protocol requires homogeneous microwave power across the imaged area, and thus suffers from a reduced field of view compared to the CW QDM; however, this limitation is more than compensated by measurement speed.  The 125$\times$125 $\mu$m$^2$ field-of-view reported in \cite{DQ4R}, with a 1 $\mu$m extent in z and 1 $\mu$m lateral resolution, implies 64,000 one-second exposures or 18 hours of measurement time to completely image a mm$^3$ detector segment.  

Unlike the CW QDM experiments in Table \ref{table:Samples}, the magnetic sensitivity reported in \cite{DQ4R} is not directly analogous to strain sensitivity – the measurement protocol used is, in fact, insensitive to crystal strain.  Pulsed-microwave protocols, constructed to be sensitive to strain and temperature variations and unaffected by magnetic field gradients, have been used for sensitive NV thermometry \cite{LukinThermometry2013,WrachtrupThermometry2013,AwschalomThermometry2013}.  As discussed in Sec.~\ref{sec:Stateoftheart}, temperature and strain both induce common-mode shifts in NV transition frequencies, and spatially resolved measurements of those shifts measure strain.  The same pulse sequences used for NV thermometry, if implemented in an imaging configuration similar to \cite{DQ4R}, would be appropriate to measure strain and should resolve averaging time issues with QDM strain spectroscopy.  In fact, strain-sensitive pulse sequences promise to improve sensitivity beyond the equivalent magnetic-field sensitivity demonstrated in \cite{DQ4R} - insensitivity to magnetic noise from impurity spins can dramatically increase coherence times \cite{HodgesTimekeeping2013}.  This direction will be pursued in future work.

The other outstanding challenge to NV strain spectroscopy is restricting the sensed region in $z$, or depth into the diamond.  Even high numerical aperture optics with small depth of field collect out-of-focus light, reducing sensitivity to small strains.  A number of optical sectioning methods commonly used in biological imaging could be adapted to imaging with NV centers.  In particular, structured illumination microscopy (SIM) takes advantage of the increased attenuation away from the focal plane of light modulated with high spatial frequency to reject out-of-focus, unmodulated light \cite{ChaslesOpticsExpress} and has been implemented to achieve superresolution imaging in diamond \cite{SIMNVs2014}.  Light-sheet microscopy uses a cylindrical lens to create a thin sheet of laser light, restricting the excitation volume \cite{WholeBrainLightSheet}; this technique has also been implemented in diamond to measure the properties of microwave devices \cite{NVLightSheet}.  Implementation in diamond of these techniques must be improved to enable $\mu$m localization of nuclear recoil tracks - SIM has so far only been used in diamond to achieve lateral superresolution with a quasi-two-dimensional distribution of NV centers, while the thinnest light sheet in \cite{NVLightSheet} was 14 $\mu$m.  Nevertheless, given extensive development of these techniques in the biosciences, achieving a cubic $\mu$m voxel is a realistic target.

An alternative technique, scanning confocal laser microscopy, inherently rejects out-of-focus light.   The confocal excitation and detection spot can be scanned through the sample volume, giving three-dimensional information (as shown in Fig.~\ref{fig:confocalstrain} below).  A confocal setup allows improved collection optics and MW power delivery, which could improve sensitivity compared to the imaging configuration by an order of magnitude or less, giving measurement times $>$10 ms per scan point.  Measuring $\sim$10$^9$ voxels of 1 $\mu$m$^3$ volume is therefore challenging; for comparison, each point would need to be measured in under 100 $\mu$s to scan one mm$^3$ in one day.  Improved wide-field strain spectroscopy therefore presents a more realistic approach.

\subsection{Pre-existing strain in CVD diamond}
\label{sec:strainbackground}

Measuring small strains quickly is only one component of detecting the signature from WIMP recoil tracks. Additionally, the host diamond must exhibit sufficiently low strain to make such measurements possible.  Pre-existing strain can interfere with damage track detection in two ways.  First, large strain gradients can degrade the performance of a QDM, both by shifting the NV spin resonance away from the microwave drive frequency and by degrading the coherence properties of individual NV centers \cite{StrainPaper}.  Second, intrinsic strain features with the right length and size could appear to be damage tracks themselves, yielding false positives.  

\begin{figure}[htbp!]
\begin{center}
\includegraphics*[width=\columnwidth]{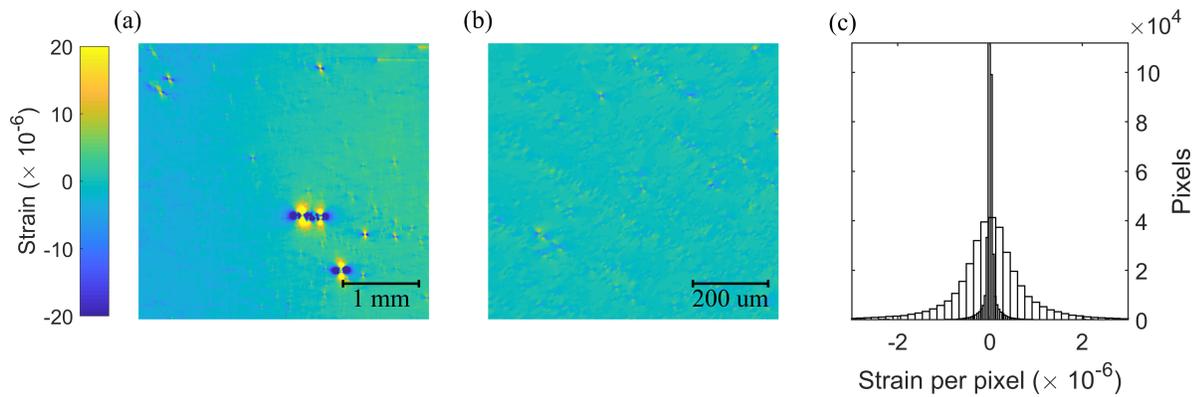}
\caption{Quantum diamond microscope (QDM) strain images of a 40 $\mu$m-thick, NV-rich diamond layer grown on an electronic-grade single-crystal diamond substrate.  \textbf{(a)} Wide-field QDM image reveals minimal strain at the millimeter scale across the sample, allowing strain-sensitive measurements without retuning or parameter changes.  Also visible are several spatially large, high-strain features propagating from imperfections at the substrate surface.  \textbf{(b)} Increasing the magnification and subtracting the linear gradient observed in (a) reveals smaller, weaker strain features.  \textbf{(c)} A histogram of strain per pixel in (b), before (white bars) and after (black bars) applying a Gaussian high-pass spatial filter.  Note that the uncertainty of these measurements is approximately 1$\times10^{-7}$ -- the distribution represents the range of strains present in the diamond.  The standard deviation before high-pass filtering, approximately 1$\times10^{-6}$, is comparable to the range of strain signal amplitudes predicted in Table \ref{table:strains}. After high-pass filtering, the standard deviation of $1.5\times10^{-7}$ is below all but the weakest predicted signal amplitudes.  Carefully evaluation of feature geometry would improve this discrimination further, and allow identification of WIMP-induced damage tracks.}
\label{fig:QDMHound}
\end{center}
\end{figure}

All CVD diamond is grown by deposition of diamond crystal layers on a pre-existing substrate, and recent advances in processing of these substrates --- as well as other growth techniques --- allow creation of NV-rich bulk diamond with low strain inhomogeneity \cite{PurpleDiamond,Friel2009}.  These processes suppress both large-scale gradients and high-gradient pixels.  Initial measurements show that state-of-the-art sensing diamonds already have sufficiently controlled wide-field strain to enable a DM search.  A benchmark for suitably suppressed large-scale strain gradients would be an NV shift below 1 MHz - approximately equal to fractional strain of 3$\times10^{-5}$.  Keeping large-scale strain gradients below this level would allow all NV centers to be addressed by the same MW drive field (assuming a realistic Rabi frequency $\geq$10 MHz in a pulsed-MW measurement).  Figures \ref{fig:QDMHound}a and b show QDM strain maps of representative NV-rich CVD diamonds, grown using strain-relieving techniques, which meet this criterion.

\begin{figure}[htbp!]
\begin{center}
\includegraphics*[width=.9\columnwidth]{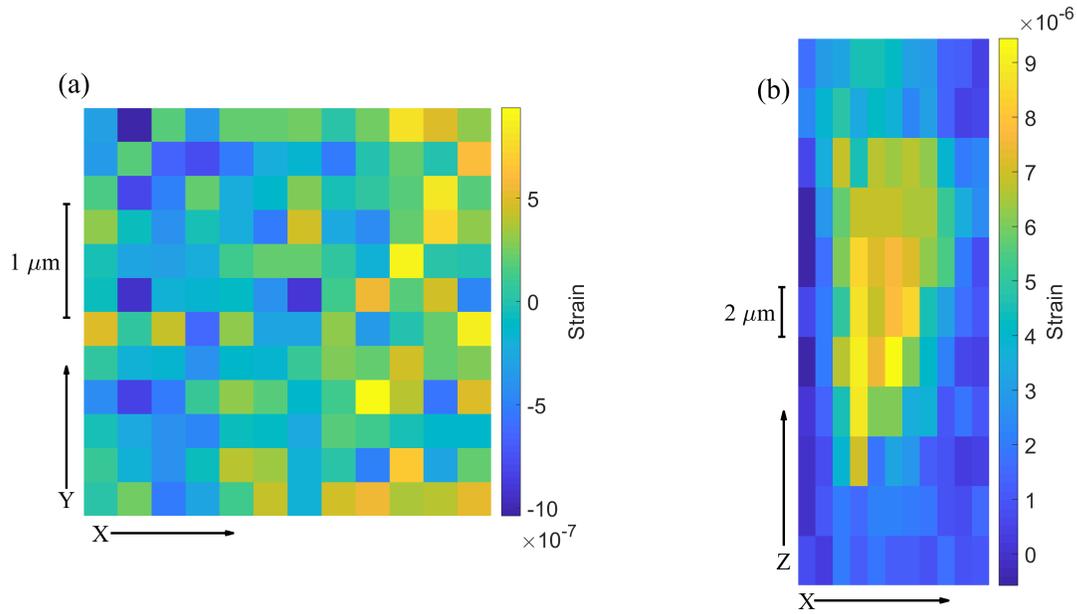}
\caption{Scanning confocal microscope NV strain maps.  \textbf{(a)} High-resolution, narrow-field survey of a 500 $\mu$m-thick, NV-enriched bulk diamond.  Standard deviation of strain per pixel of 4$\times$10$^{-7}$ approaches levels needed for WIMP-induced damage track detection.  \textbf{(b)} Strain measurement along the crystal growth (axial, ``Z'') direction of one of the small strain features shown in Fig.\ \ref{fig:QDMHound}b.  Growth defect strain features can be distinguished from damage tracks by their axial extent.}
\label{fig:confocalstrain}
\end{center}
\end{figure}

While this 1 MHz benchmark for large-scale strain gradients is larger than the expected WIMP-induced strain signal magnitude, nuclear recoil track signals can be distinguished from large-scale growth defects via their spatial geometry -- any feature that extends more than a micron in any direction can be assumed not to result from a WIMP impact.  Figures \ref{fig:QDMHound}a and b demonstrate strain features typical for CVD-grown NV-rich diamond thin layers, which nucleate at the substrate interface and propagate upwards, forming a characteristic ``petal'' feature \cite{StrainPaper}.   See Fig.~\ref{fig:confocalstrain}b for a measurement of strain in one such feature along a slice parallel to the growth direction. Features such as those shown in Fig.~\ref{fig:QDMHound} would therefore not prevent detection of a nearby WIMP track.  We can approximate such geometric discrimination by applying a spatial high-pass filter to QDM data.  Figure \ref{fig:QDMHound}c shows the distribution of strain in individual pixels of Fig.~\ref{fig:QDMHound}b, both with and without such spatial filtering.  Without spatial filtering, the strain level in intrinsic strain features is comparable to the expected strain in a WIMP track; after high-pass filtering, the remaining strain features are smaller than all but the weakest predicted WIMP-induced strain signals.

In addition to impeding wide-field NV strain spectroscopy, pre-existing strain could also impede a WIMP search if it created a risk of false positive damage track detections.  Pre-exposure scanning of the entire large-volume ($\sim$m$^3$) detector would be prohibitively time-consuming.  The prevalence of intrinsic strain features that mimic damage tracks must therefore be suppressed below 1/mm$^3$ in diamond material suitable for a directional WIMP detector.  Such features must either be forbidden by the growth process or removed by post-processing of the detector material.  Our initial measurements, as well as previous studies of defects in diamond \cite{Friel2009,PintoExtendedDefects,PurpleDiamond,MakiVacancyAnnealing2007,BangertVacancies2009}, indicate that this criterion is fulfilled by strain-relieving growth conditions combined with post-growth high-temperature annealing of diamond. However, further characterization of large volumes of material will be required once strain imaging techniques have been developed that meet the criteria discussed in Sec.~\ref{sec:improvingqdm}.

There are two types of possible strain-inducing features at relevant length scales for false positives (larger than point defects, but smaller than a micron): linear crystal growth defects and extended vacancy clusters \cite{Friel2009,PintoExtendedDefects}.  Crystal growth defects can be suppressed by careful surface processing \cite{Friel2009} and choice of diamond growth parameters.  Additionally, while these defects may be under a micron wide, they propagate upwards from the substrate during growth; this propagation renders them easily distinguishable from recoil tracks if the 1 $\mu$m slice for imaging is chosen to be perpendicular to the growth direction \cite{E6xray2008}, or by combining axial slices to measure the three-dimensional shape of the feature, as in Fig.\ \ref{fig:confocalstrain}b.  Meanwhile, high-pressure, high-temperature (HPHT) treatment of diamond has been found to eliminate nanoscale vacancy clusters, causing the vacancies to aggregate into large platelets containing $\sim$10$^4$ vacancies, which can be easily distinguished from WIMP-induced damage tracks \cite{MakiVacancyAnnealing2007,BangertVacancies2009}.  Initial high-resolution x-ray diffraction measurements over limited diamond volumes show no evidence of strain-inducing features with length scales between 20 and 100 nm \cite{ourXray}, indicating that they either are not produced during deposition or are broken up during annealing.  Confocal scanning strain measurements have also not found evidence of submicron-scale strain features in bulk NV-rich diamond (for example, see Fig.~\ref{fig:confocalstrain}a).  We note that these measurements were performed over limited sample volumes, and only show that these features are not widespread.  Two benchmarks for development of a prototype detector would be a background search covering $>$1 mm$^3$ of diamond at $\mu$m resolution, as well as experimental characterization of diamond processing by measuring ion-induced submicron strain features \cite{AusStrain2019} before and after annealing and pressure treatment.

In summary, QDMs have demonstrated a diamond strain measurement noise floor equivalent to the 1$\times10^{-7}$ required to detect WIMP-induced damage tracks.  We estimate a pulsed-microwave protocol will image a mm$^3$ detector segment in 18 hours, fast enough to enable operation of a m$^3$-scale detector.  A pulsed-microwave protocol should also have an improved noise floor compared to that demonstrated with past QDM measurements, due to insensitivity to the magnetic spin bath and faster measurement times (which leave less scope for thermal fluctuations).  Major milestones still to be met include demonstrating such sensitivity in a strain-sensing modality; implementing optical sectioning methods to achieve $\mu$m$^3$ resolution; and, once those are achieved, background studies demonstrating suppression of pre-existing false-positive strain features below 1/mm$^3$ via careful growth and post-growth processing.

\subsection{Cryogenic zero phonon line NV spectroscopy}
\label{sec:cryoZPL}

In addition to shifting the microwave transition frequency of the NV electron's spin, local strain also shifts the optical transition frequency of the NV orbital energy level \cite{OpticalStressTensor2013}.  This frequency shift could  be used to identify high-strain pixels in a  damage track search.  Doing so requires spectroscopy on the NV zero phonon line (ZPL) - the electronic energy level transition without absorption or emission of an accompanying phonon.  At cryogenic temperatures, strain-induced shifts in the ZPL frequency have been used to precisely locate NVs within a diffraction-limited spot: e.g., relative positions of nearby NV centers in polycrystalline diamond were resolved to about 1 nm using a tunable laser \cite{EnglundStrainZPL}.  (At room temperature the ZPL is too broad to measure small strain shifts --- for temperatures below 100 K, the ZPL linewidth decreases proportionally to T$^5$ because of the dynamic Jahn-Teller effect \cite{dynamicjahnteller2009}.)  For wide-field track localization, the detector segment under investigation could be held at cryogenic temperatures, and the ZPL interrogated with such a tunable laser.  NV centers under sufficient local strain will fluoresce at a distinct excitation wavelength from their neighbors; NV centers near a WIMP-induced damage track should experience strains well over $10^{-6}$, corresponding to shifts in the ZPL of order 1 GHz \cite{OpticalStressTensor2013}.  Detecting such wavelength-dependent excitation in widefield imaging would be challenging with traditional cameras due to low photon counts; however, electron-multiplying (EM) CCD arrays have been used for widefield imaging of fluorescence from single NV centers \cite{EnglundStrain2016, NVPLE2012} with sufficient sensitivity for real-time position tracking of nanodiamond diffusion \cite{LevitatedNanodiamond2013}.  Such a device could measure wavelength-dependent photoluminescence in the scheme described here.

Measuring GHz shifts in optical frequency would require only modest laser stabilization. However, to be measurable by wavelength-dependent excitation, the $\sim$GHz-scale shift must be larger than the linewidth of the ZPL.  A wide range of values have been measured for the transition linewidth using single NVs at cryogenic temperatures, from  as low as 10 MHz with NVs in bulk diamond \cite{StarkShiftControl}, to 1 GHz with NVs in thin diamond membranes \cite{MembranePRX2017}, and up to tens of GHz for NV centers within nanometers of a diamond surface \cite{CuiThesis2014}.  To our knowledge, no such measurements have been performed for NVs in an ensemble sufficiently dense to enable WIMP track detection; performing such a measurement would be a first step in implementing this technique.  

Assuming the ZPL for NVs in an ensemble is sufficiently narrow, the measurement time required for this method will depend on the camera framerate and number of frames per field of view (which is equivalent to the number of excitation wavelengths measured).  For example, measuring a range of $\pm1$ GHz in 10 MHz steps at a framerate of $\sim$60 FPS \cite{cameralimits} would require approximately three seconds per FOV.  (An equivalent FOV to the 125$\times$125 $\mu$m$^2$ demonstrated for NV spectroscopy is consistent with demonstrated capabilities of EM-CCD cameras and NV collection optics.)  This is slower than the estimated one second per FOV for optimized strain spectroscopy, but still could meet the 1-3 day/mm$^3$ benchmark to outpace the expected neutrino floor background event rate.  

\section{Damage track localization with NV center creation}
\label{sec:NVFNTD}

The crystal damage from a WIMP candidate track is expected to include dozens to hundreds of recoil-induced lattice site vacancies.  Vacancies in diamond become mobile under high-temperature annealing, and may combine with a nearby nitrogen impurity to form an NV center \cite{JelezkoNVFormation2014,FuNVFormation2020}.  In a diamond with high nitrogen content, recoil-induced vacancies could be annealed to produce NV centers at the site of a WIMP event.  If the high-nitrogen diamond also had few initial NVs, this would enable a nearly background-free localization via the fluorescence of the induced NV centers.  

Fluorescence microscopy on radiation-induced defects as a method for nuclear recoil detection is a mature technique, largely using Al$_2$O$_3$:C,Mg crystals referred to as ``fluorescent nuclear track detectors'' (FNTDs) \cite{FNTDsoriginal2006}. These provide an instructive case study for track localization via induced fluorescence. Color centers --- F$_2^{2+}$(2Mg) --- are formed in Al$_2$O$_3$:C,Mg during crystal growth.  Nuclear recoils induce local electron cascades, transforming the color centers via electron capture to F$_2^+$(2Mg) centers, which exhibit a distinct absorption and emission spectrum.  Confocal microscopy combined with reconstructive algorithms yield 3D reconstruction of the nuclear trajectory \cite{FNTD3D2013}, which has been applied to radiation oncology/dosimetry \cite{HybridCellFNTD2013,FNTDAlphaDosimetry2018}, as well as nuclear \cite{FNTDNuclearCharge2014}, neutron \cite{FNTDNeutrons2018}, and beam \cite{FNTDSimulation2017} physics.

FNTD crystal studies have focused either on energy ranges well above 10--100 keV, or long tracks generated via irradiation with neutrons.  In both cases, tracks are typically much longer than the optical diffraction length, and can be measured using confocal microscopy.  Superresolution methods such as STED \cite{FNTDSTED2017} and SIM \cite{AlphaFNTDwithSIM2018} have been applied in some cases, but the best reported resolution is large compared to the length scale of an expected WIMP damage track.

A similar principle has been applied in diamond --- recoil tracks are mapped via creation of NV centers using radiation-induced vacancies \cite{DiamondFNTD2015}.  After exposure to high-energy ions, the diamond is annealed at high temperature, allowing the vacancies to migrate.  Vacancies that migrate near pre-existing nitrogen impurities are captured, forming NV centers; these are detected via confocal fluorescence microscopy.  Unlike the FNTD crystals, this process requires high-temperature annealing to ``develop'' the nuclear recoil tracks --- recoil-induced damage in diamond is then used to study vacancy diffusion during annealing \cite{DiamondFNTD2017}.  However, a key difference between FNTD crystals and diamonds is the much lower fluorescence per unit of deposited energy in diamond, due to low conversion yield of vacancies to NV centers.

\begin{figure}[htbp!]
\begin{center}
\includegraphics*[width=\columnwidth]{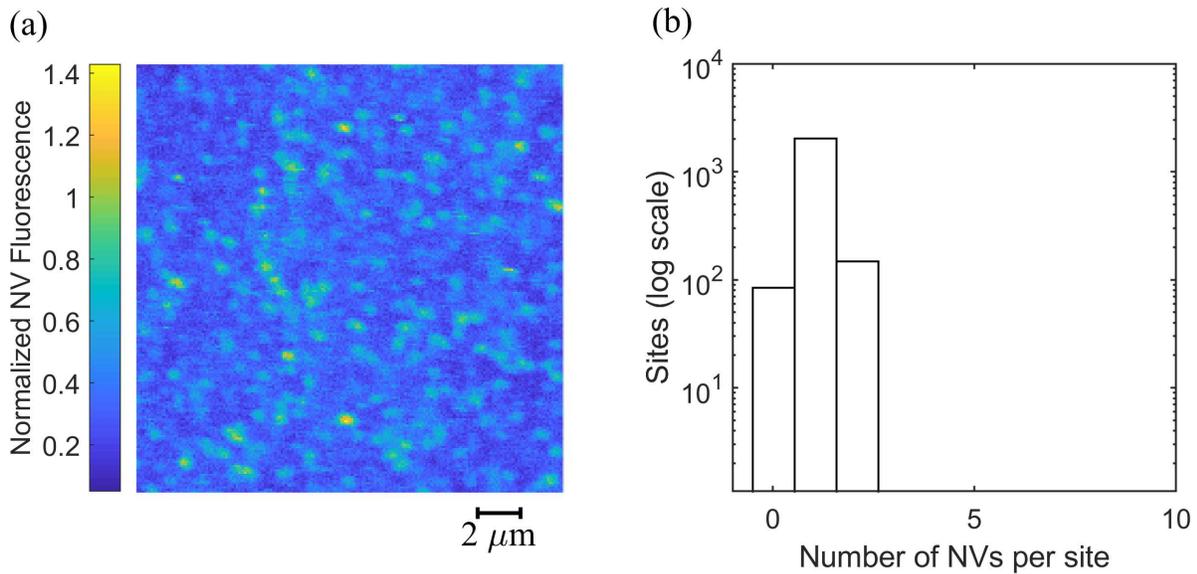}
\caption{Confocal fluorescence collection on an annealed, high-[N], low-[NV] diamond suitable for fluorescent nuclear track detector (FNTD) studies.  \textbf{(a)} Typical 10$\times$10 $\mu$m$^2$ field of view demonstrating the pre-existing single-NV background.   \textbf{(b)} NV center positions were counted by identification of local maxima in confocal fluorescence scans of bulk diamond samples.  The intensity of fluorescence at each site was then used to determine the number of NVs in a diffraction-limited spot.  Local maxima categorized as ``0'' NV centers represent out-of-focus emitters, where the confocal microscope collects fluorescence below half the rate of an in-focus NV center.   A limited initial survey found no spots with more than 2 NV centers per diffraction-limited volume in $\sim10^4$ $\mu$m$^3$ of three diamond samples; see text for details.}
\label{fig:FNTDbackground}
\end{center}
\end{figure}

Previous fluorescent track detection research in diamond has used long damage tracks from high-energy particles --- higher by a factor of $10^4$ than relevant energies for WIMP detection \cite{DiamondFNTD2015,DiamondFNTD2017}.  With many fewer than $10^4$ NVs created per ion track, this method would be unable to detect a WIMP-induced recoil.  However, a modified version of this protocol using a high-nitrogen diamond could increase sensitivity to a level capable of WIMP track detection.  Type Ib diamonds, grown via the high-temperature, high-pressure (HPHT) process, feature nitrogen impurities at hundreds of parts per million, together with very few vacancies or pre-existing NV centers \cite{ChenHPHTvsN2018, NirNVTEM2017}.  Initial studies, as shown in Fig.\ \ref{fig:FNTDbackground}, find that while there are features near the surface that could be construed as false positives, no such features are found in the bulk of the diamond, indicating they are not inherent to the growth process and could be eliminated with advanced surface treatment \cite{PurpleDiamond}.  Initial measurements, examples of which are presented in Fig.~\ref{fig:FNTDbackground}, find no more than two NV centers in any diffraction-limited volume.  This brief study examined three diamond samples over a total volume of $\sim$10$^4 \mu$m$^3$, and identified $\sim$3,000 NV centers.  The study was performed with a scanning confocal microscope focused 5-6 $\mu$m below the diamond surface, to avoid polishing defects and surface imperfections, and characterized the intrinsic emitter distribution of the diamond material.  Assuming similar results from future studies of larger volumes, this method would give essentially background-free detection of nuclear recoil tracks so long as they yield 3 or more NV centers upon annealing.  Because of this low background, this method could allow fast confocal scanning microscopy using the $\sim$100 $\mu$s dwell time required to scan a mm$^3$ detector segment in one day; alternatively, the fluorescence could be quickly imaged in a wide-field modality, using a low-light camera such as an EM-CCD.  

To estimate whether low-energy WIMP collisions will lead to a detectable number of created NVs, we perform simulations of recoil cascades using the TRIM software package \cite{SRIM}.  These simulations indicate that a 10 keV recoil induces several dozen vacancies extending over approximately 20 nm.  The migration range of these vacancies during annealing can be controlled by choice of temperature and annealing time. To preserve head-tail asymmetry \cite{proposal}, vacancies must migrate much less than 20 nm during the annealing process.  Assuming, for our estimate, a 3 nm vacancy migration length and a typical nitrogen impurity concentration of 200 ppm for a type Ib HPHT diamond, there will be $\sim$20 N nuclei available within the vacancy travel range for NV center formation during annealing.  The actual number of NVs created is difficult to predict, as it will depend not only on the number of vacancies and nitrogen nuclei available, but also on recombination rates with interstitial carbon and formation of vacancy clusters \cite{JelezkoNVFormation2014,NVFormationTheory2014,HaqueNVFormation2017}.  A detailed experimental measurement of the NV yield as a function of nuclear recoil energy and annealing parameters is necessary to demonstrate feasibility of this technique.  

If track location via induced fluorescence proves viable, the high nitrogen concentration will complicate nanoscale track mapping via superresolution spectroscopy (see section \ref{sec:superres}) by reducing the NV center coherence and dephasing times \cite{TakahashiSpinCalcs2013,AkimovHPHTNVSpins2018}.  If the vacancy-to-NV conversion rate is high, the track direction may be determined by resolving the positions of the individual induced NVs; with lower yields, non-NV-based methods may be required, such as the X-ray spectroscopy described in section \ref{sec:xray}.

In summary, our small initial background study of high-nitrogen type Ib diamonds found sufficiently low NV densities to allow background-free detection of nuclear recoil tracks.  Calculations and SRIM simulations indicate sufficient numbers of both nitrogen impurities and recoil-induced vacancies near the damage site to create as many as 20 NV centers, even for the lowest-energy WIMP events.  A key measurement going forward will be to determine the NV yield as a function of recoil energy using implanted-ion test signals; the lowest energy recoil that reliably creates 3 or more new NV centers will be a minimum energy threshold for the technique.  If this proves viable, a follow-up study will investigate the effect of high-temperature annealing on damage track structure and directionality.  

\section{Nanoscale track direction mapping}
\label{sec:superres}
Once the position of a WIMP (or neutrino) induced damage track has been determined to within a micron-scale voxel, the incident direction of the initial particle must be extracted from the spatial distribution of crystal damage. Damage tracks exhibit head-tail asymmetry and orientation correlated with the direction of the WIMP candidate \cite{proposal}.  In this section, we will discuss methods to map this crystal damage at the nanoscale.  Extracting the direction from a track as short as 20--50 nm (for 10--30 keV initial recoil energy) requires a three-dimensional measurement with resolution below the track length.  

\subsection{Superresolution NV strain spectroscopy}

Rapid developments in NV-based magnetic field sensing with spatial resolution below the optical diffraction limit have been motivated by the pursuit of nanoscale MRI \cite{NVNanoReview2014,NVMRIreview2016,NVMRIreview2019}.  As with wide-field magnetic imaging, techniques developed for subdiffraction magnetic sensing with NV centers can be adapted to sense crystal strain with a change of microwave pulse sequence. For WIMP track mapping, a superresolution technique must be capable of three-dimensional operation with $\sim20$ nm resolution, and simultaneously allow microwave spectroscopy of the NV spin.  Broadly, three classes of superresolution techniques have been applied to NV centers: deactivation techniques involving a doughnut-shaped laser beam, stochastic techniques for separating individual NV centers, and techniques based on strong magnetic field gradients.  We will review each of these and discuss applications to directional WIMP detection.  

Several deactivation techniques have been used to localize individual NV centers separated by less than the diffraction limit or perform superresolution spectroscopy.  In each case, the signal from NV centers in a doughnut-shaped laser beam spot is suppressed, allowing information to be collected only from the NVs at the small central intensity minimum of the doughnut beam.  Stimulated emission depletion, or STED, uses a strong pulsed doughnut beam to induce stimulated emission from the addressed NVs, depleting their fluorescence before data is collected.  In diamonds with a thin layer of NV centers or in nanodiamonds, STED has been used for NV center localization to $\sim5$ nm \cite{EarlySTEDHell2009} and magnetic spectroscopy with spatial resolution $<10$ nm \cite{SILSTED2012,STEDnanodiamonds2013}.  Charge state depletion (CSD) replaces the strong depletion beam with a doughnut beam at a wavelength chosen to photoionize NV$^-$ centers to NV$^0$; in combination with optical filtering blocking NV$^0$ fluorescence, this allows interrogation only of the remaining NV$^-$ at the doughnut beam's central minimum.  CSD microscopy has been demonstrated to localize nearby NV centers with 4 nm resolution \cite{CSDmicroscopy2015}, and to perform magnetic superresolution magnetic spectroscopy in a thin, dense layer of NV centers with 45 nm spatial resolution \cite{CSDSuperresMagImg2019}.  Finally, spin-RESOLFT imaging uses a doughnut beam after a magnetic sensing microwave pulse sequence, destroying sensing information stored in NV centers under the doughnut beam and allowing spectroscopy only on NV centers in the central minimum \cite{EarlySpinRESOLFT2010}.  Using shallow NV centers, this enabled magnetic spectroscopy with 20 nm spatial resolution \cite{SpinRESOLFT2017}.  

Past demonstrations of these techniques have used either quasi-two-dimensional ensembles or relatively sparse single NV centers, which do not overlap in the axial direction.  To measure a damage track, which could be anywhere in the diamond segment, requires three-dimensional resolution.  A number of methods for generation of three-dimensional doughnut-hole beams in diamond have been explored; using a mirror and optical interference \cite{MirrorSTED2016} or patterned light \cite{3DSTEDHell2009} yield $\sim100$ nm resolution in the axial dimension.  In biological media, two crossed doughnut beams have been used to perform STED microscopy with 45 nm resolution in all three dimensions \cite{nondiamond3DSTED2008}, although applying this technique in diamond is complicated by the high index of refraction.  Therefore, we expect these techniques to be very challenging to perform in three dimensions at the sub 20 nm resolution required to infer damage track direction.

Stochastic superresolution imaging methods, such as STORM and PALM, are widely used in biosciences and have been demonstrated using NV centers, achieving 12 nm localization \cite{EnglundSTORM2013} and magnetic sensing with 20 nm spatial resolution \cite{STORMWrachtrup2014}.  While high NV densities impede stochastic localization of individual emitters \cite{SpinRESOLFT2017}, these techniques may prove useful in HPHT diamonds coupled with vacancy creation by damage tracks described in section \ref{sec:NVFNTD}. 

An alternative method for superresolution spectroscopy uses magnetic gradients to obtain spatial localization from the NV spin precession frequency rather than the optical transition.  Large magnetic gradients can be applied so that only a small class of NV centers resonate with a uniformly applied microwave control sequence, allowing spatially selective imaging with high resolution using magnetic gradient coils \cite{FourierSelectiveEnsemble2017} or magnetic write heads \cite{HardDriveWrachtrup2018}.  Microfabricated gradient coils enabled magnetic spectroscopy on selectively addressed NV centers with 30 nm resolution \cite{FourierSelectiveWalsworth2017}.  Magnetic field gradients can also be used for position encoding, rather than position-selective addressing.  In Fourier magnetic imaging, the gradient encodes an NV's position onto its spin precession phase and frequency; post-processing the acquired images enables subdiffraction imaging.  This method has been used for magnetic field imaging with a two-dimensional set of single NVs with 30 nm spatial resolution \cite{AraiFourierImaging2015}.  A similar method using microwave field inhomogeneity as a localization tool was demonstrated for 2D magnetic imaging with 100 nm spatial resolution in a dense NV ensemble \cite{WidefieldNanoMRIWrachtrup2019}.  

2D spatial resolution at or near the 20 nm level has been demonstrated with all of the superresolution methods described above. This benchmark is required for determining damage track directionality.  These methods are also compatible with strain-sensitive microwave interrogation.  The major outstanding challenge with any of these techniques will be to employ them in a dense, three-dimensional ensemble of NV centers.  With the exception of recent charge state depletion \cite{CSDSuperresMagImg2019} and microwave-inhomogeneity assisted localization \cite{WidefieldNanoMRIWrachtrup2019}, demonstrations of superresolution NV spectroscopy have typically been performed in relatively low density samples, with only a few NV centers per diffraction-limited volume; a conservative estimate of the NV density required for track mapping is  several hundred in such a volume \cite{proposal}. Meanwhile, 3D superresolution NV imaging has only been demonstrated with STED, and only with 100 nm resolution \cite{MirrorSTED2016,3DSTEDHell2009} (although similar techniques could be used for CSD or spin-RESOLFT imaging).  To extract the direction of a 10--30 keV WIMP event will require significant improvements in optics, either by improved spatial light modulation or by successful application of crossed-beam STED in a diamond sample.  Fourier magnetic imaging could also reach the required resolution, but generating appropriately large, structured three-dimensional gradients using fabricated microcoils remains challenging.  

\begin{figure}[htbp!]
\begin{center}
\includegraphics*[width=\columnwidth]{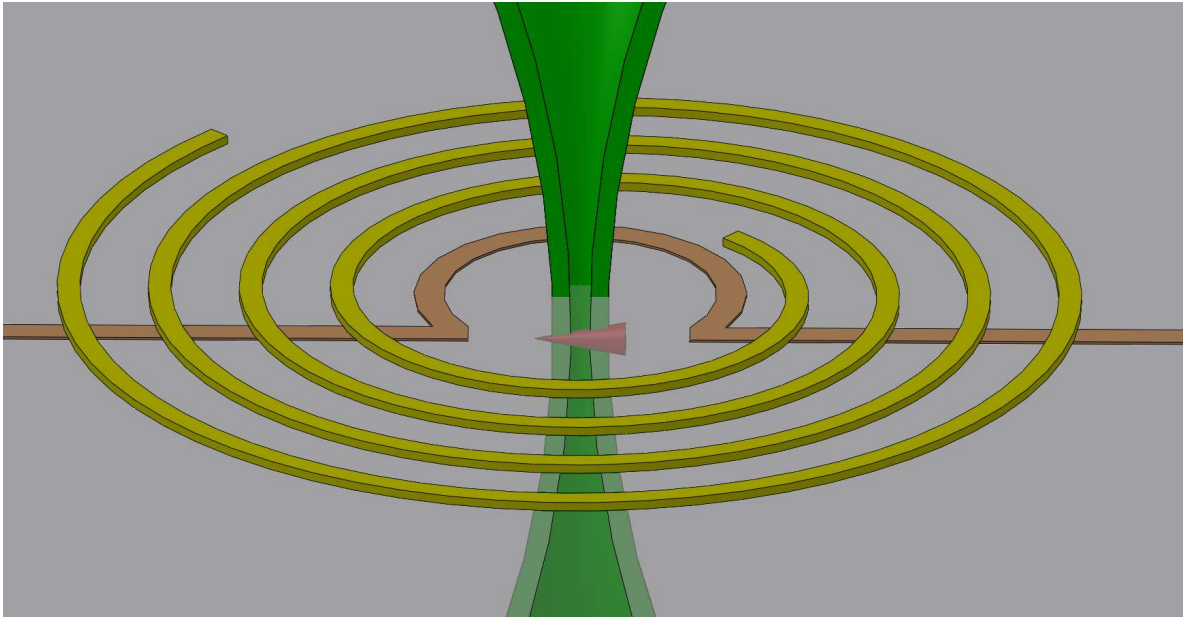}
\caption{Proposed combined technique for three-dimensional superresolution strain spectroscopy.  A doughnut optical beam, implementing either STED or CSD, enables lateral superresolution by depleting surrounding NVs.  A microcoil creates a magnetic field gradient, which allows Fourier deconvolution along the vertical axis.  The microcoil and microwave guide are fabricated around the micron-localized damage track position.}
\label{fig:doughnut-coil}
\end{center}
\end{figure}

Given the current state of the art, a combination of methods, as illustrated in Fig.~\ref{fig:doughnut-coil}, will likely be required to measure the directionality of damage tracks. A two-dimensional STED or CSD beam could provide $<$20 nm resolution in the lateral plane, while a magnetic field gradient enables selective addressing or post-processing deconvolution for $<$20 nm axial resolution.  A doughnut beam would be created using standard STED optics, and an appropriate microfabricated gradient coil would provide one-dimensional deconvolution of the axial position.  We can approximate the time required for such a measurement from the number of samples needed to attain 10 nm axial resolution, and the averaging time per sample used in previous Fourier deconvolution experiments.  The combined method illustrated in Fig.~\ref{fig:doughnut-coil} would require acquisition of Fourier-encoded data at $100\times100$ 10-nm beam spots to cover the entire 1 $\mu$m voxel within which the damage track is pinpointed.  To achieve 10 nm axial resolution within a 1 um spot (the approximate axial extent of a confocal STED or CSD beam spot), Nyquist sampling implies $\sim$200 samples.  Taken together with the data acquisition time of $\sim$20 ms per sample required in prior demonstrations of Fourier magnetic imaging with NV centers \cite{AraiFourierImaging2015}, this gives a time estimate of $\sim$11 hours to image the $\mu$m voxel in three dimensions with 10 nm resolution -- well within the target established by the neutrino floor event rate.

The optical superresolution methods described in the previous paragraphs may not be applicable to the high-nitrogen, low-NV case described in section \ref{sec:NVFNTD}.  Magnetic field noise from the dense bath of nitrogen impurity spins reduces the NV coherence time, complicating methods like spin-RESOLFT and Fourier magnetic imaging, which require precise spectroscopy or quantum state control.  Conversely, the low density of pre-existing NV centers means that a damage track's direction could be mapped simply by localizing the created NV centers, without requiring strain spectroscopy.  

One further caveat is worth mentioning - the sub-diffraction experiments and superresolution benchmarks described above were all performed using NV centers at or near the surface of the diamond, whereas a WIMP candidate event can happen anywhere within the mm$^3$ detector section.  Deeper into the diamond, the optical point spread function of any microscope is both broadened and distorted by diamond's high refractive index \cite{HaeberlePSFs2003}.  However, once the damage track is localized at the um$^3$ scale within the detector segment, diamond can be etched away until the damage track is within a few $\mu$m of the nearest surface.  Using appropriate techniques \cite{Friel2009}, this should be possible without introducing additional damage or strain and without distorting the WIMP signal.  (We note that the low-strain etching technique described in \cite{Friel2009} removes only 40-90 nm/min \cite{LeeEtching2008}, resulting in average etch times of O(20) hours to reach damage tracks buried in the diamond.  Performing this etching process in parallel with a second, faster process would keep the etching from being a rate-limiting step. For example, a recently reported wet etching process on diamond achieved removal rates as high as 9 $\mu$m/min \cite{NagaiEtching2018}.  Combinations of etching techniques have been shown to achieve fast etch rates on diamond while eliminating etch-induced flaws \cite{HicksEtching2019}.)

In summary, three-dimensional mapping of WIMP-induced damage track structure using superresolution strain spectroscopy of NV centers is achievable but would require some advances in NV measurements.  The requisite sub-20 nm resolution has been achieved in two-dimensional or sparse collections of NV centers using a variety of methods, some of which have also been demonstrated using dense ensembles.  The key experimental benchmark for this approach will be demonstrating high-resolution, three-dimensional spectroscopy of NV centers; a hybrid approach incorporating both optical depletion and Fourier deconvolution methods gives a realistic path to achieve this goal while keeping measurement times accordant with the predicted neutrino floor event rate.

\subsection{Nanoscale x-ray strain measurement}
\label{sec:xray}

Strain measurements can also be made with spatial resolution below the diffraction limit for visible light by using shorter-wavelength photons.  In particular, hard x-ray photons have subnanometer wavelengths and can be focused to spot sizes below 10 nm \cite{sub10nmxray2010}.  Scanning x-ray diffraction microscopy (SXDM) presents an alternative method for nanoscale mapping of damage tracks.  A variety of techniques exist for high-spatial-resolution imaging of materials structure with x-ray nanobeams, with spatial resolution of 2--50 nm and strain resolution at the 10$^{-4}$ to 10$^{-5}$ level \cite{NanobeamReview2018,HoltReview2013}.  Hard x-ray nanobeams with appropriate resolution and sensitivity are currently found only at a few synchrotron facilities worldwide \cite{APSNanoprobe2012,NSLSNanoprobe2017,NanoMAXnanoprobe2018}; however, for the low expected annual event rate \cite{proposal} and with accurate optical localization, the damage track direction for all candidate WIMP events could be read out with a modest amount of x-ray beam time per year. 

SXDM makes use of a highly focused beam spot, scanned across the surface of a sample held at the Bragg condition \cite{HoltReview2013}; at each beam position, a segmented detector records a diffraction pattern.  This diffraction pattern is sensitive to small perturbations of the crystal structure within the beam spot, allowing strain and crystal orientation to be mapped with high spatial resolution \cite{nonptychonanobeam2011}.  Strain measurements with spatial resolution as low as 5 nm \cite{ThinFilmCDI2013} have been performed using coherent Bragg ptychography - diffraction patterns from overlapping coherent beam spots can be inverted by an iterative algorithm, giving high-resolution images of the diffracting crystal structure \cite{BraggCDI2014}.  Additionally, three-dimensional images can be retrieved by coherent Bragg ptychography, while maintaining resolution below the beam spot size \cite{3DBPP2017}.  

While SXDM combines high strain and spatial resolution, a diffraction pattern at any given Bragg angle will only be sensitive to a particular subset of strain vectors - those that change the crystal lattice spacing for the plane from which the x-ray photons diffract.  To fully map a damage track, it would therefore be necessary to scan the crystal from multiple Bragg angles. In addition to covering a full range of strain vectors, measuring at multiple angles would allow for three-dimensional reconstruction without ptychography; additionally, multiple-angle Bragg ptychography allows improved three-dimensional imaging, relaxing the demanding experimental constraints of single-angle Bragg ptychography \cite{multiangleBPP2018}.

Appropriate spatial resolution in three dimensions has been demonstrated with SXDM.  Strain resolution near 10$^{-5}$ is routinely attainable; while the strain averaged over a voxel at the visible-light diffraction limit is as low as 10$^{-6}$-10$^{-7}$, within 100 nm of the track it will be well over 10$^{-5}$. The x-ray beam is focused only in two dimensions - the focal length is limited only by absorption and diffraction in the diamond, and can extend hundreds of times the lateral spot size.  While the shift in average strain across the entire beam spot due to a damage track may be below the detection threshold, the strained volume of diamond will diffract some of the beam to an observable angle. It should thus be possible to measure a small feature with large strain by looking for the photons diffracted from the strained subvolume rather than a shift in the average strain over the entire beam spot.  This principle has been demonstrated in diamond in preliminary studies for WIMP track measurement \cite{ourXray}.  Features with length scales smaller than that of a damage track have been detected with SXDM - for example, atomic-scale variations in thickness of a 10 nm Si quantum well were measured in an Si/SiGe heterostructure \cite{SiQuantumWells2012}.  Measurement times with SXDM are typically of order one second per beam position; measuring a $\mu$m$^2$ field of view in 10 nm steps would therefore take approximately three hours.  The need to acquire data at multiple projection angles will increase this time, but it will always be below the target set by the neutrino floor event rate -- fully defining the strain tensor requires at most six projection angles \cite{3DStrainCDI2010}, and full three-dimensional imaging is possible with fewer angles using advanced reconstruction techniques \cite{3DBPP2017,multiangleBPP2018}.

As with the optical superresolution methods described above, SXDM measurements would require that, before measuring a damage track's direction, the diamond would first be etched to bring the damage track within a few microns of the surface. Additionally, the detector segments would need to be appropriately shielded in transit from the detector site to the synchrotron facility against cosmic rays.  (Ideally, any etching should be performed at the same facility as superresolution mapping, allowing the surrounding diamond to help shield the cubic micron of interest during transport.)   Notably, this method would be compatible with non-spectroscopic methods of damage track localization, including the low-yield scenario discussed in section \ref{sec:NVFNTD} for track detection via NV center creation.

\section{Outlook and next steps}

The two major technical challenges facing directional WIMP detection in diamond, damage track localization and direction measurement, have solutions requiring challenging but realistic development over existing technology. We aim, in ongoing experiments, to demonstrate the requisite improvements.  We are developing two methods for damage track localization: strain spectroscopy and creation of new color centers from recoil-induced damage.  One set of improvements to strain spectroscopy will adapt existing magnetic-field imaging techniques to achieve sensitive, high-spatial-resolution strain measurements; a second experiment demonstrating restriction of the excitation light with either light-sheet or structured-illumination microscopy will follow.  Event localization via fluorescent nuclear track detection requires a 10$^4$ reduction in detected ion energy over previous experiments, but with high-nitrogen, low-fluorescence diamonds this should be possible; characterization of fluorescence creation under these conditions is underway.  

Two methods for extracting the incoming particle direction via nanoscale mapping of damage tracks are also under consideration.  Three-dimensional superresolution strain spectroscopy using a combination of a two-dimensional doughnut-shaped laser beam and one-dimensional Fourier magnetic gradient deconvolution would only require combining existing technologies, while cryogenic zero phonon line spectroscopy offers a superresolution method using only Gaussian optics.  Finally, synchrotron measurements with a nanofocused x-ray beam are also underway.  The initial step is to demonstrate separate measurement of a small, strained subvolume in an unstrained, NV-rich diamond; injected-ion measurements to evaluate sensitivity will follow.

Reaching sensitivities below the neutrino floor for WIMP dark matter searches with reasonable exposures will likely require directional detection.  Diamond is a promising material for future WIMP target development.  Building on over a decade of research into the NV center and the excellent crystalline and optical properties of CVD and HPHT diamond should enable directional detection in diamond with further technique development.  A combined high-energy/atomic physics approach may then provide directional dark matter searches with solid-state density, opening a route to WIMP sensitivity below the neutrino floor.

\section{Acknowledgements}
We acknowledge technical discussions and assistance from Pauli Kehayias, Connor Hart, Raisa Trubko, Reza Ebadi, Alex Sushkov, Martin Holt, Tao Zhou, F. Joseph Heremans, Nazar Delegan, and Surjeet Rajendran.  This work was supported by the DOE QuANTISED program under Award No. DE-SC0019396; the Army Research Laboratory MAQP program under Contract No. W911NF-19-2-0181; the DARPA DRINQS program under Grant No. D18AC00033; and the University of Maryland Quantum Technology Center.

\section{References}
\bibliographystyle{iopart-num.bst}
\bibliography{iopart-num.bib}

\end{document}